\documentclass[doublecol,final]{epl2} 
\usepackage{graphicx}              % list packages between braces
\usepackage{color}
\usepackage{amsmath}
\usepackage{amssymb}

\title{Spin-blockade qubit in a superconducting junction}
\author{C. Padurariu \inst{1} \and Yu. V. Nazarov \inst{1}}
\shortauthor{C. Padurariu \etal}

\institute{                    
  \inst{1} Kavli Institute of NanoScience, Delft University of Technology - Lorentzweg 1, 2628 CJ, Delft, The Netherlands
}

\abstract{
We interpret a recent pioneering experiment [Zgirski M. {\it et al.}, {\it Phys. Rev. Lett.}, \textbf{106} (2011) 257003] on quasiparticle manipulation in a superconducting break junction in terms of spin blockade drawing analogy with spin qubits. We propose a novel qubit design that exploits the spin state of two trapped quasiparticles. We detail the coherent control of all four spin states by resonant quantum manipulation and compute the corresponding Rabi frequencies. The read-out technique is based on the  spin-blockade that inhibits quasiparticle recombination in triplet states.   We provide extensive microscopic estimations of the parameters of our model.
}

\newcommand{\Mab}[4]{\left[\begin{array}{cc}
\hspace{-0.1cm} #1\ & \hspace{-0.3cm} #2 \hspace{-0.1cm}\\
\hspace{-0.1cm} #3\ & \hspace{-0.3cm} #4 \hspace{-0.1cm} \hspace{-0.1cm}\end{array} \right]}

\newcommand{\twovec}[2]{\left[\begin{array}{r}
\hspace{-0.1cm} #1 \hspace{-0.1cm}\\
\hspace{-0.1cm} #2 \hspace{-0.1cm}\end{array} \right]}

\newcommand{\ket}[1]{\left\vert#1\right\rangle}
\newcommand{\bra}[1]{\left\langle#1\right\vert}

\newcommand{\projector}[2]{\left\vert#1\right\rangle\left\langle#2\right\vert}
\newcommand{\sandwich}[3]{\left< #1 \vphantom{#2 #3} \right| #2 \left|\vphantom{#1 #2} #3 \right>}

\begin{document}

\maketitle

\section{Introduction}
The spin degree of freedom provides a natural representation of quantum information inspiring some of the first qubit designs~\cite{LossDiVincenzo}. Nanodevices that realize read-out and manipulation of single electron spins are in the focus of modern research~\cite{single}.
 An alternative is provided by Josephson-based superconducting qubits~\cite{SupQubitsReview} where the qubit states emerge from the interplay of Josephson effect and Coulomb blockade, not having anything to do with real spin. There are theoretical proposals~\cite{NazarovSSQ, SSQpaper} attempting to combine the advantages of both qubit realizations. This can be achieved by a superconducting spin qubit, where quantum information is stored in the state of a spin trapped in a superconducting junction. 

Generally, in this situation one deals with the spin of a superconducting quasiparticle rather than with an electron spin. The supercondcuting quasiparticles are excitations of the Bardeen-Cooper-Schrieffer (BCS) superconducting ground state that are spin doublets despite the fact that they do not bear a definite charge, being superpositions of electron and hole excitations~\cite{BCSoriginal}.

One needs to localize a quasiparticle to control its spin state. It is possible to trap a quasiparticle  not only inside a superconducting island~\cite{SupIslands}, but also in eventually any  superconducting junction. The quasiparticles are kept in  the Andreev bound states (ABS) that develop~\cite{AndreevDots} in the presence of a superconducting phase difference dropping at the junction. The presence of an additional quasiparticle in a bound state is manifested by a change of the superconducting current. In addition, spin-orbit interaction in the junction makes the superconducting current sensitive to the spin state of the trapped quasiparticle~\cite{NazarovSSQ}. This provides a way to detect the quasiparticle occupancy, as well as to manipulate the spin state. Using ABS is particularly advantageous since in this case  the junction may support larger superconducting current as compared to the systems involving Coulomb islands~\cite{Ruitenbeek}.

Controllable trapping of quasiparticles in ABS and their detection has proven an experimental challenge and has been realized only recently~\cite{MainExperiment}. The experiment uses a mechanically controlled break of Al strip to produce atomic-size junction of adjustable  transparency. The junction normal state conductance is contributed by a few transport channels; their transparencies are accurately characterized from the fit of I-V curves. In the superconducting state, each channel gives rise to an ABS that can accommodate quasiparticles.  

Detection is based on the fact that the critical current of the junction is altered by the presence of trapped quasiparticles. The measurement of critical current must be achieved with a time resolution smaller than the quasiparticle escape time. The experiment uses rectangular pulses of bias current of typical duration $\simeq 1$ $\mu$s and different amplitudes. Each pulse induces with a certain probability the switching of the junction from the zero-voltage to running state. The latter is manifested as a measurable voltage response. The probability is determined by repeating the measurement a big number of times $\simeq 10^4$ for each value of the pulse amplitude~\cite{MainExperiment}. Upon increasing  the pulse amplitude the switching probability increases in steps. The step positions indicate the critical current of the junction with different numbers of trapped quasiparticles and are in good agreement with the theoretical predictions. In addition, the measurement provides information about the relative probability of different occupancies of the ABS. These probabilities are proportional to the increase in switching probability at the corresponding step.

The following experimental detail has inspired the proposal presented in this Letter. In most cases either zero or one quasiparticles (at lowest level) have been detected.  However, the state with two trapped quasiparticles has been reported as well. It could not be observed in a junction with a single high-transparency channel, but has been observed for junctions supporting two open channels. Importantly, the measured critical current corresponded to the case when the quasiparticles are situated in different levels: the lowest one and the next-to lowest one. 

In this Letter, we interpret this detail as the {\it spin-blockade} of two quasiparticles. Spin-blockade in a double quantum dot connected to two leads~\cite{SpinBlockadeFirst,Petta,Oleg} is utilized for the read-out and operation of the most practical spin qubits. 
The transport cycle through the dots is arranged in such a way that the state with two electrons in different dots can only transit to a spin-singlet state with two electrons in the same level in one of the dots. Spin conservation forbids transitions between singlet and triplet. Therefore the electron transport is blocked if two electrons in two different dots happen to be in a triplet state.
  
Similarly, two quasiparticles in two different Andreev levels : lowest and next-to-lowest --- form the ground state of the system with full spin $S=1$. Spin conservation in this case forbids a transition to $S=0$ no-quasiparticle state of the lower energy, the transition proceeds quickly if the two-quasiparticle state is a singlet. One can say the junction is blocked in a triplet state. This guarantees long life-time and therefore observability of these specific two-quasiparticle states.

 We show that this spin blockade in combination with spin manipulation techniques can be used to control and read out a simple qubit. We develop a minimum theoretical model that includes all relevant spin-dependent phenomena in the junction and provide extensive microscopic estimations for the parameters of the model. We detail the resonant quantum manipulation of all four spin states formed by two quasiparticles and outline a "natural" read-out in the system. %and provide estimation of  quantum coherence of the qubit
Experimental realization of our proposals would for the first time unambiguously prove the spin properties of superconducting quasiparticles and open up new ways to combine spin and superconductivity in the context of quantum technologies.

\section{Theoretical model}

\begin{figure}
\onefigure[width=0.8\linewidth]{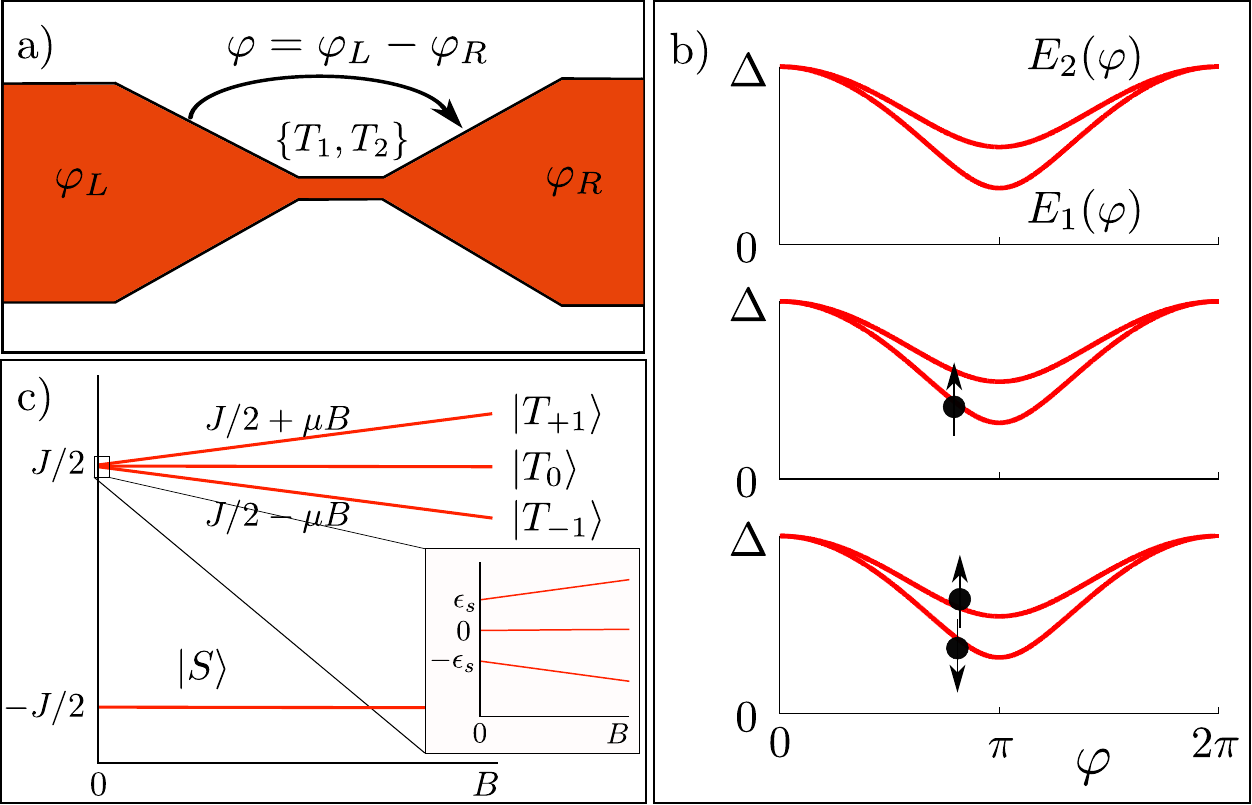}
\caption{Andreev bound states in a junction with two open channels. \textbf{a)} Sketch of the junction characterized by the channel transmission coefficients $\{T_1,T_2\}$ and the superconducting phase difference $\varphi$. \textbf{b)} Energy of Andreev states as a function of $\varphi$. The three plots show most long-lived states stable with respect to spin-conserving energy relaxation. 
%These occupancy states can be observed experimentally in the case of a two-channel junction; 
They are: the state with no quasiparticles  ($S=0$), the state with a single quasiparticle ($S=1/2$), and the state with two quasiparticles in a triplet spin configuration ($S=1$). \textbf{c)} The energies of the four spin states of two quasiparticles plotted versus the spin magnetic field. Inset: Zoom in the low $B$-field region to show the spin-orbit splitting of the triplet states. The energy of the spin-orbit splitting is small compared to the exchange splitting $\epsilon_{\rm SO}\ll J$.}
\label{fig:spectrum}
\end{figure}

The break junction is characterized by a set of transport channels labelled by $p$  with transmission coefficients $0<T_p<1$. The Andreev levels corresponding to the channels have energies $E_p(\varphi) = \Delta (1-T_p \sin^2(\varphi/2))^{1/2}$,  where $\Delta$ is the bulk superconducting gap ($\simeq 1.9\times 10^{-4}$ $eV$ in~\cite{MainExperiment}). Each level may accommodate up to two quasiparticles of opposite spin. We introduce the level occupation $n_{p}=0,1,2$. 

The occupation of the Andreev levels  determines the phase-dependent part of the junction energy, which is given by the sum over the levels,
\begin{equation}
E(\varphi) = \displaystyle \sum_{p} (n_p-1) E_p(\varphi), 
\end{equation} 
and the superconducting current flowing in the junction,
\begin{equation}
I_s= \frac{2e}{\hbar}\frac{\partial E(\varphi)}{\partial \varphi} =\frac{2e}{\hbar} \displaystyle \sum_{p} (n_p-1) \frac{\partial E_p(\varphi)}{\partial \varphi} .
\end{equation} 
We focus on the case when two {\it lowest} levels are occupied with a single quasiparticle each, $n_1=n_2=1$, $n_p =0$ if $p \ne 1,2$. We call these states TQTL (two quasiparticles, two levels). In this case, the two lowest levels do not contribute to the current resulting in a suppression of its magnitude. Two spin-1/2 quasiparticles in a TQTL give rise to four spin states: a singlet $\ket{S}$ and three triplet states $\ket{T_j}$, $j=-1,0,+1$. 

Since the TQTL is an excited state , one needs to drive the junction out of equilibrium to realize it. In~\cite{MainExperiment} this has been achieved by applying a current pulse of large amplitude just before each measurement of the critical current. The current in the pulse exceeds a critical one and results in quasiparticle generation at the junction. After the pulse, some of the quasiparticles generated are trapped in the ABS. Two quasiparticles are trapped with some probability $P_2$. The quasiparticles can relax their energy by going to the lower Andreev levels. If their state is a singlet, they may also annihilate: we assume that this process is fast at the scale of the measurement time. The annihilation takes place also in the case of a singlet TQTL: in this case, we characterize it with the rate $\Gamma_d$.  If the quasiparticles are in the triplet state, they relax to TQTL and stay there manifesting themselves in a lower critical current. It is reasonable to assume random initial spin: in this case, the probability to measure the lower current is $P = (3/4) P_2$. This is provided that the measurement time is shorter than the spin-orbit relaxation rate of the triplet state to be estimated below.

In addition to the decay of the singlet state, the evolution of the TQTL is determined by three spin-dependent effects: the exchange interaction, the spin-orbit (SO) coupling and the interaction with an external magnetic field $\vec{B}$. We summarize these effects in a Hamiltonian $H_s$.
\begin{equation}
\label{eq:H_s}
H_s=H_{\rm exchange}+H_{\rm SO}+H_{\rm B}
\end{equation}

The exchange interaction term is  expressed as usual in terms of the two spin operators $\vec{\sigma}_1$ and $\vec{\sigma}_2$ at the levels $1,2$, $H_{\rm exchange} = J(\vec{\sigma}_1 \cdot \vec{\sigma}_2)$,  $J$ being the exchange coupling. It is convenient to rewrite the term in the basis of singlet and triplet states.
\begin{equation}
\label{eq:H_exchange}
H_{\rm exchange}=-J/2 \ket{S}\bra{S}+J/2\sum_j \ket{T_j}\bra{T_j}
\end{equation}

The effect of SO interaction on a spin state of a quasiparticle occupying an Andreev level has been discussed previously in~\cite{NazarovSSQ,SSQpaper}. The interaction is characterized by a level-specific pseudo-vector $\vec{\epsilon}_p$ of dimension energy that polarizes the qusiparticle spin. 

We denote the SO pseudo-vectors of the two Andreev levels involved by $\vec{\epsilon}_1$ and $\vec{\epsilon}_2$. The vectors generally depend on the superconducting phase drop $\varphi$ over the junction. The following symmetry holds: $\vec{\epsilon}_{1,2}(\varphi)=-\vec{\epsilon}_{1,2}(-\varphi)$. It is a consequence of time-reversal symmetry preserved by the SO interaction. The SO Hamiltonian takes the form~\cite{SSQpaper} $H_{\rm SO}=\sum_{n=1,2}\vec{\epsilon}_n(\varphi)\cdot\vec{\sigma}_n$. To express it in singlet-triplet basis it is convenient to introduce the sum and difference of pseudo-vectors $\vec{\epsilon}_{s,d}=(\vec{\epsilon}_1\pm\vec{\epsilon}_2)/2$, following  \cite{Oleg}.
\begin{align}
\label{eq:H_SO}
H_{\rm SO}= & \sum_{j}j\epsilon^z_s \ket{T_{j}}\bra{T_{j}}+\sum_{\pm}\frac{\epsilon^x_s\pm i\epsilon^y_s}{\sqrt{2}} \ket{T_{0}}\bra{T_{\pm 1}}+\\
\ & \epsilon^z_d\ket{S}\bra{T_0}+\sum_{\pm}\frac{\mp \epsilon^x_d- i\epsilon^y_d}{\sqrt{2}} \ket{S}\bra{T_{\pm 1}} +h.c.\notag
\end{align}
The spin-quantization axis $z$ is not yet fixed here. We will make a particular choice of the quantization axis further in the text. 

We also take into account the spin interaction with an external magnetic field $H_{\rm B}=\mu \sum_{n=1,2} \vec{B}\cdot\vec{\sigma}_n$. Here, the magneton $\mu$ is smaller than its value in vacuum. The reason for this is that the Andreev bound states may spread into superconducting leads at distance exceeding the screening length of the magnetic field. In what follows it is convenient to set $\mu \vec{B} \to \vec{B}$, introducing a field $\vec{B}$ of dimension energy. In the singlet-triplet basis $H_{\rm B}$ takes the form
\begin{align}
\label{eq:H_B}
H_{\rm B}=& \sum_{j}jB^z\ket{T_{j}}\bra{T_{j}}+ 
\\
\ & \sum_{\pm}\frac{B^x\pm iB^y}{\sqrt{2}} \ket{T_{0}}\bra{T_{\pm 1}} +h.c.\notag
\end{align}

The magnetic field dependence of the energies of the resulting states is given in Fig.~\ref{fig:spectrum}c.

The dynamical evolution of the four spin states is adequately described by a density matrix equation which takes into account the unitary evolution described by $H_s$, as well as the decay of the singlet component with the rate $\Gamma_d$. The density matrix $\rho$ is defined in the space of four spin states of interest, $\rho_{xy}=\sandwich{x}{\rho}{y}$, $x,y\in \{S,T_{-1},T_0,T_{+1}\}$. Owing to the decay, ${\rm Tr} \rho < 1$. The equation set reads:
\begin{align}
\frac{d}{dt}\rho=&\ i[H_s,\rho] - \Gamma_d \rho_{SS} \projector{S}{S}-\label{eq:general_eom}\\
\ &\ \sum_{j}\left(\frac{\Gamma_d}{2}\rho_{ST_j}\projector{S}{T_j}+h.c.\right)\notag
\end{align}
There are four  equations to solve for diagonal components , as well as six for off-diagonal ones. It is relatively easy to obtain the numerical solution for arbitrary values of the parameters. To reveal the physics of the spin system we present analytical solutions in certain parameter regimes relevant for experiments similar to~\cite{MainExperiment}. 

Before doing this, we need to estimate the parameters of our model to reveal the relevant time scales. 

\section{Estimations}

\begin{figure}
\onefigure[width=0.6\linewidth]{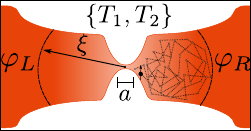}
\caption{Sketch of the atomic-size break junction. The wavefunction of quasiparticles trapped in localized Andreev states extends away from the atomic-size constriction of length $a$ to a much longer distance of the order of the superconducting coherence length $\xi$. Assuming the dirty-limit for the Al film superconductor used in the experiment, the motion of the quasiparticle inside the volume taken by the state follows a chaotic scattering trajectory.}
\label{fig:junctionlength}
\end{figure}

Let us start with the energy relaxation mechanisms in the junction. These mechanisms may involve the transitions of quasiparticles to lower Andreev levels, as well as their annihilation. 
The relaxation process dumps energy into an environment and thus must be accompanied by an emission of a photon, or of a phonon. Since the states are localized inside the superconductor where electromagnetic field is screened, the phonon mechanism dominates.

For the junction setup in~\cite{MainExperiment} the situation is depicted in Fig.~\ref{fig:junctionlength}. The wavefunction of a  localized quasiparticle extends in the superconducting leads for a length of the order of the superconducting coherence length, $\xi$. Assuming the dirty-limit for the Al film superconductor~\cite{dirty}, the coherence length can be estimated as $\xi\simeq (\hbar D/\Delta)^{1/2} $
%\simeq 8.3 \times 10^{-8}$ $m$, 
where $D$
%\simeq 2\times 10^{-3}$ $m^2/s$ 
is the diffusion coefficient of electrons in normal metal. The atomically-thin constriction extends for a length $a\simeq 1..10 \times 10^{-10}$ m, therefore $a\ll \xi$. For the film geometry, the volume taken by the wavefunction of the ABS can be estimated as ${\cal V}\simeq \xi^2 d$, $d$
%\simeq 1.0\times 10^{-7}$ $m$ 
being the film thickness. 

Since the volume ${\cal V}$ is large at atomic scale, the rate can be estimated from the rate of phonon-mediated energy relaxation for normal-metal electron states forming a continuous spectrum. This rate at energy $E$ is given by $ \Gamma_{r}(E) = \kappa_{\rm el-ph} E^3$, $\kappa_{\rm el-ph}$ being the electron-phonon coupling constant ($\kappa_{\rm el-ph}\simeq 3\ \mu {\rm s}^{-1}{\rm K}^{-3}$~\cite{el-ph} for Al.) For our estimation, we need the transition rate $\Gamma_d$ between two discrete states. To get this, we divide $\Gamma_r$ by number of states $N$ available for relaxation, $N \simeq \nu E {\cal V}$, $\nu$ being the electron density of states. Therefore, $\Gamma_d \simeq\Gamma_r /N$. To adjust the estimation to the ABS case, we estimate $E$ by $\Delta$ and ${\cal V}$ by the volume taken by an ABS.

If we express $\xi$ in terms of diffusion coefficient, the estimation of $N$ appears strikingly simple, $N \simeq (R_\Box G_Q)^{-1}$, $G_Q \equiv \pi e^2/\hbar$, $R_\Box$ being the normal-state film resistance per square. One can argue that the same estimation holds for an arbitrary geometry provided $R_\Box$ is replaced with an effective resistance $R_{\rm eff}$ of a metal piece of size $\simeq \xi$ adjacent to the constriction.  This is a manifestation of the famous Thouless relation~\cite{bookNazarov} between level spacing and dwell time in a piece of disordered metal.

With this, we estimate the ABS relaxation rate as
\begin{align}
\Gamma_d \simeq (G_Q R_{\rm eff})\Gamma_r(\Delta) \simeq \kappa_{\rm el-ph} \Delta^3 G_Q R_{\rm eff} \ .
\label{eq:singlet-decay} 
\end{align}

This estimation of the phonon relaxation rate shall be compared with the results of Ref. \cite{Ivanov} where $\Gamma_d \simeq \Gamma_r(\Delta)$ was obtained.
Since Ref. \cite{Ivanov} assumed one-dimensional geometry, the effective (Sharvin) resistance of the setup can be estimated as $1/G_Q$. This demonstrates consistency of our results. It has been already argued in \cite{Zazunov} that the rate of Ref. \cite{Ivanov} is suppressed by a factor accounting for the lateral spread of the ABS wavefunction in the leads. We have thus shown that the suppression factor is conveniently expressed in terms of the effective resistance $R_{\text{eff}}$.

Further, let us estimate the exchange coupling  $J$ as the interaction energy of two electrons (or holes) localized in the volume ${\cal V}$. Owing to the strong screening in the bulk of the metal, the interaction quenches if the distance between the electrons exceeds the atomic scale $a$. The probability for electrons to be that close is estimated as $a^3/{\cal V}$.  This yields  $J\simeq E_{\text{at}}(a^3/{\cal V})$, where we estimated the interaction by the atomic energy scale $E_{\text{at}}$.
 Since $\nu \simeq (E_{\text{at}} a^3)^{-1}$  the estimation of $J$ can be expressed in terms of the number $N$ introduced above,
\begin{align}
J \simeq \frac{E_{\rm at} a^3}{{\cal V}}\simeq \frac{\Delta}{N}\simeq \Delta (R_{\rm eff}G_Q)\ .
\label{eq:exchange}
\end{align}
%Using the values of parameters specified we find $J\simeq 1.9 \times 10^{-9}$ $e\text{V}$. 
We compare our estimations of $\Gamma_d$ and $J$ to find that exchange splitting exceeds by far the life-time broadening of the singlet state,  $\hbar\Gamma_d/J \simeq \Delta^2 \hbar \kappa_{el-ph} \simeq 10^{-4}$. We note that the ratio is independent of the effective resistance $R_{\rm eff}$ and is thus rather universal for all ABS setups.

The last estimation we need is that of the SO splitting $\simeq |\vec{\epsilon}_{\rm SO}|$. In the case of a general superconducting constriction of size of the order of $\xi$, the scale of SO splitting is given by $\alpha^{\rm SO}_{\rm Al} \Delta$ \cite{NazarovSSQ}, where the SO coefficient in Al  $\alpha^{\rm SO}_{\rm Al}\simeq 10^{-2}$ can be interpreted as a probability to flip the electron spin in course of a scattering event. This however implies the energy dependence of the transmission amplitudes at a scale of $E \simeq \Delta$. For a short break junction, this energy dependence is absent; this cancels the SO splitting \cite{NazarovSSQ}. The energy-dependent part of the transmission amplitude comes about the scattering of an electron that has passed the constriction, proceeded at distance of the order of $\xi$ in a lead and got back to the constriction (see Fig.~\ref{fig:junctionlength}). The estimation for SO splitting is thus reduced by the backscattering probability. The latter can be again estimated as $1/N \simeq R_{\rm{eff}} G_Q$. Thus, we find
\begin{align}
|\vec{\epsilon}_{\rm SO}| \simeq \alpha^{\rm SO}_{\rm Al} \Delta/N(\Delta) \simeq \alpha^{\rm SO}_{\rm Al} \Delta R_{\rm eff}G_Q \ ,
\label{eq:SO}
\end{align}
%leading to $|\vec{\epsilon}_{\rm SO}| \simeq 1.9 \times 10^{-13}$ $eV$. 
We compare spin and exchange splitting to find the former to be a factor of $|\vec{\epsilon}_{\rm SO}|/J\simeq \alpha^{\rm SO}_{\rm Al}\simeq 10^{-2}$ smaller not depending on the setup details.

Let us provide typical scales of the estimated parameters. For an experiment with $R_{\rm eff}\simeq 10$ Ohm (that is, $N\simeq 10^3$) and $\Delta/\hbar \simeq 10^{11}$ Hz the above estimations yield $J\simeq 10^8$ Hz, $|\vec{\epsilon}_{\rm SO}|\simeq 10^6$ Hz and $\Gamma_d\simeq 10^4$ Hz. We refer to these values in following concrete estimations.

\section{Spin-orbit relaxation of the triplet states}

We proceed to describe the dynamics of the triplet TQTL states in conditions of time-independent magnetic field and superconducting phase drop at the junction. Quasiparticle annihilation in the triplet states is forbidden by spin blockade. However, the SO interaction given in eq.~(\ref{eq:H_SO}) provides a relaxation mechanism by coupling the triplet states with the singlet. The coupling is described by the difference of SO pseudo-vectors $\vec{\epsilon}_d$.

We compute the SO relaxation rate by treating the small quantities $\epsilon_d/J\simeq \alpha^{\rm SO}_{\rm Al}$ and $\hbar\Gamma_d/J$ as perturbations. It is convenient to choose the spin quantization axis along the direction of vector $(\vec{B}+\vec{\epsilon}_s)$ and work with the parallel (perpendicular) component of $\vec{\epsilon}_d$ with respect to this direction, $\epsilon^{\parallel}_d$ ($\epsilon^{\perp}_d$). We find the decay rates $\Gamma_{T_j}$ corresponding to the triplet states $\ket{T_j}$, $j=-1,0,+1$.
\begin{equation}
\Gamma_{T_{\pm 1}}=\Gamma_d \left(\epsilon^{\perp}_d\right)^2/(J \pm B)^2,\quad \Gamma_{T_{0}}=\Gamma_d \left(\epsilon^{\parallel}_d\right)^2/J^2.
\end{equation}
We estimate the decay rates $\Gamma_{T_j} \simeq \Gamma_d(\alpha^{\rm SO}_{\rm Al})^2$ yielding $\Gamma_{T_j} \simeq 1$ Hz. We assume that the triplet states are long-lived at the scale of the measurement time.

\section{Singlet-to-triplet manipulation}

We detail here the resonant manipulation of the singlet-to-triplet transition. 
The pseudovector $\vec{\epsilon}_d$ that defines the couplings between the singlet and triplet states depends on  
the superconducting phase $\varphi$ 
that can be modulated by an electric signal at frequency $\Omega$,
\begin{equation}
\varphi(t)=\varphi_0+A\left[\exp(i\Omega t)+\exp(-i\Omega t)\right]
\label{eq:delta_phi}
\end{equation}
$A$ being the dimensionless modulation amplitude $A\ll 1$. The corresponding modulation of the pseudovector is then given by:
\begin{equation}
\vec{\epsilon}_{d}(t)=\vec{\epsilon}_{d}(\varphi_0)+\frac{d\vec{\epsilon}_{d}}{d\varphi}(\varphi_0)\ A\left[\exp(i\Omega t)+\exp(-i\Omega t)\right]\ .
\label{eq:delta_eso}
\end{equation}

The condition of resonant manipulation is achieved when $\hbar\Omega$ matches the singlet-triplet energy spacing, that is different for different triplet states provided a sufficiently large magnetic field $B 
%\lesssim J 
\gg |\vec{\epsilon}_{\rm SO}|$ is applied . In this case, each triplet state can be addressed separately.

Let us find the change of probability to be in a triplet state as a result of a manipulation pulse of duration $\tau$. We employ  rotating wave approximation (RWA) justified by $B \gg |\vec{\epsilon}_d|,\hbar \Gamma_d $. For the transition between $\ket{S}$ and $\ket{T_j}$; the resonant condition is $\hbar\Omega=J+jB$. It is convenient to introduce slowly varying states $\ket{\tilde{S}}=\exp(-iJt/2)\ket{{S}}$ and $\ket{\tilde{T}_j}=\exp[i(J/2+j B)t]\ket{{T}_j}$ to arrive at evolution equation
\begin{flalign}
& \frac{d}{dt}\tilde{\rho}=\ i[\tilde{H}_s,\tilde{\rho}]-\frac{\Gamma_d}{2}\left(\tilde{\rho}_{S{T}_j}\projector{\tilde{S}}{\tilde{T}_j}+h.c.\right)-\notag \\
& \Gamma_d \tilde{\rho}_{SS} \projector{\tilde{S}}{\tilde{S}}; \quad \tilde{H}_s=\ \hbar\bar{\Omega}_j\ket{\tilde{S}}\bra{\tilde{T}_{j}} +h.c.
\label{eq:singlet_triplet}
\end{flalign}
where the manipulation amplitudes are $\bar{\Omega}_0 = A (d\epsilon^{z}_{d}/d\varphi)$, $\bar{\Omega}_\pm= A (d\epsilon^{\pm}_{d}/d\varphi)$,
$\epsilon^{\pm} \equiv -(\pm\epsilon^{x} + i \epsilon^{y})/\sqrt{2}$.  
We assume that the manipulation starts at time $t\gg \Gamma_d^{-1}$ after the preparation, so that ${\rho}_{SS}(0)=0$. If the initial probability to be in the triplet state is $P_j(0)$, the final one is given by
\begin{flalign}
 &P_j(\tau)=P_j(0)\left[(2\gamma^2-1)\cosh\left( \frac{\sqrt{\gamma^2-1}}{2\gamma}\Gamma_d \tau\right)+ \right.\notag\\
 &\left. 2\gamma\sqrt{\gamma^2-1}\sinh\left(\frac{\sqrt{\gamma^2-1}}{2\gamma}\Gamma_d \tau \right)-1\right]\frac{e^{-\Gamma_d \tau/2}}{2(\gamma^2-1)}
\label{eq:solution_ST}
\end{flalign}
The duration dependence is determined by the inverse dimensionless strength of the pulse $\gamma \equiv \Gamma_d/4\bar{\Omega}_j$. For weak pulses  $\gamma>1$, the singlet state decays faster than the coherent transition takes place and the triplet probability decays with the rate $\Gamma_d(1-\sqrt{1-\gamma^{-1}})$ (Fig.~\ref{fig:initialization}a). At $\gamma\gg 1$, this decay rate is twice the manipulation amplitude. For sufficiently strong pulses $\gamma<1$ one sees coherent oscillations of the probability (Fig.~\ref{fig:initialization}b) with frequency $2\bar{\Omega}_j(1-\gamma^2)^{1/2}$ on the background of the overall decay at rate $\Gamma_d$. The full depletion of $P_j(\tau)$ can be achieved by tuning the pulse duration.

\section{Triplet-to-triplet manipulation}

Resonant manipulation of the triplet-to-triplet transition is used to implement single qubit rotations. The coupling between $\ket{T_0}$ and either of $\ket{T_{\pm}}$ is realized both by SO interaction, $\vec{\epsilon}_s$, and by the magnetic field, $\vec{B}$, while the triplets $\ket{T_{+1}}$ and $\ket{T_{-1}}$ do not mix. The manipulation can be achieved by modulating either coupling parameter. However, it is practical to modulate the magnetic field, as it can be much larger than the SO splitting in the junction. 

We consider a magnetic field with a static and  an a.c. component oscillating at frequency $\Omega$
\begin{equation}
\vec{B}(t)= B \vec{z} + \vec{\tilde{B}} \exp(i\Omega t)+\vec{\tilde{B}}^*\exp(-i\Omega t)
\end{equation}
assuming $B \gg \vec{\tilde{B}}$.
The resonance condition is $\Omega=B/\hbar$, the same  for both $\ket{T_{+1}}$ and $\ket{T_{-1}}$. In RWA, the effective Hamiltonian reads ($\tilde{B} \equiv \tilde{B}_x -i \tilde{B}_y$)
\begin{align}
\tilde{H}_s=& |\tilde{B}| \ket{T_{0}}\bra{T_{s}} + h.c.;\\
\ket{T_{s}}\equiv& \frac{1}{\sqrt{2}|\tilde{B}|}\left[\tilde{B}\ket{T_{+1}} + \tilde{B}^*\ket{T_{-1}}\right]
\label{eq:triplet_manipulation}
\end{align}
describing the Rabi oscillations with frequency $\Omega_R=|\tilde{B}|/\hbar$ between $\ket{T_0}$ and a superposition state $\ket{T_s}$. After a manipulation pulse of duration $\tau$ the amplitudes $\alpha_{0}$ and $\alpha_{s}$ are transformed as 
\begin{equation}
\twovec{\alpha_{0}(\tau)}{\alpha_{s}(\tau)}= \Mab{\cos(\Omega_Rt)}{-i\sin(\Omega_Rt)}{-i\sin(\Omega_Rt)}{\cos(\Omega_Rt)}\twovec{\alpha_{0}(0)}{\alpha_{s}(0)}
\end{equation}
Assuming the static magnetic splitting is of order $B\simeq J$ and the modulation is $10\%$ of $B$, we estimate the Rabi frequency $\Omega_R\simeq 10^7$ Hz: this allows a fast and efficient triplet-to-triplet manipulation. The time available for the manipulation is set by the rate of the triplet decay $\Gamma_{T_j} \simeq 1 Hz$ so that the number of rotations can be as large as $\Omega_R/\Gamma_{T_j}\simeq 10^{7}$.

\section{Spin qubit}

Having understood the manipulation of the spin states, we can describe the design and operation of a simple spin qubit. The qubit states are $\ket{0}\equiv \ket{{T}_{0}}$ and $\ket{1}\equiv\ket{T_{s}}$. These together with the state defined below form an orthonormal basis of the triplet subspace.
\begin{equation}
\ket{2}\equiv \frac{1}{\sqrt{2}|\tilde{B}|}\left[\tilde{B}\ket{T_{+1}} - \tilde{B}^*\ket{T_{-1}}\right].
\end{equation}

\begin{figure}
\onefigure[width=0.8\linewidth]{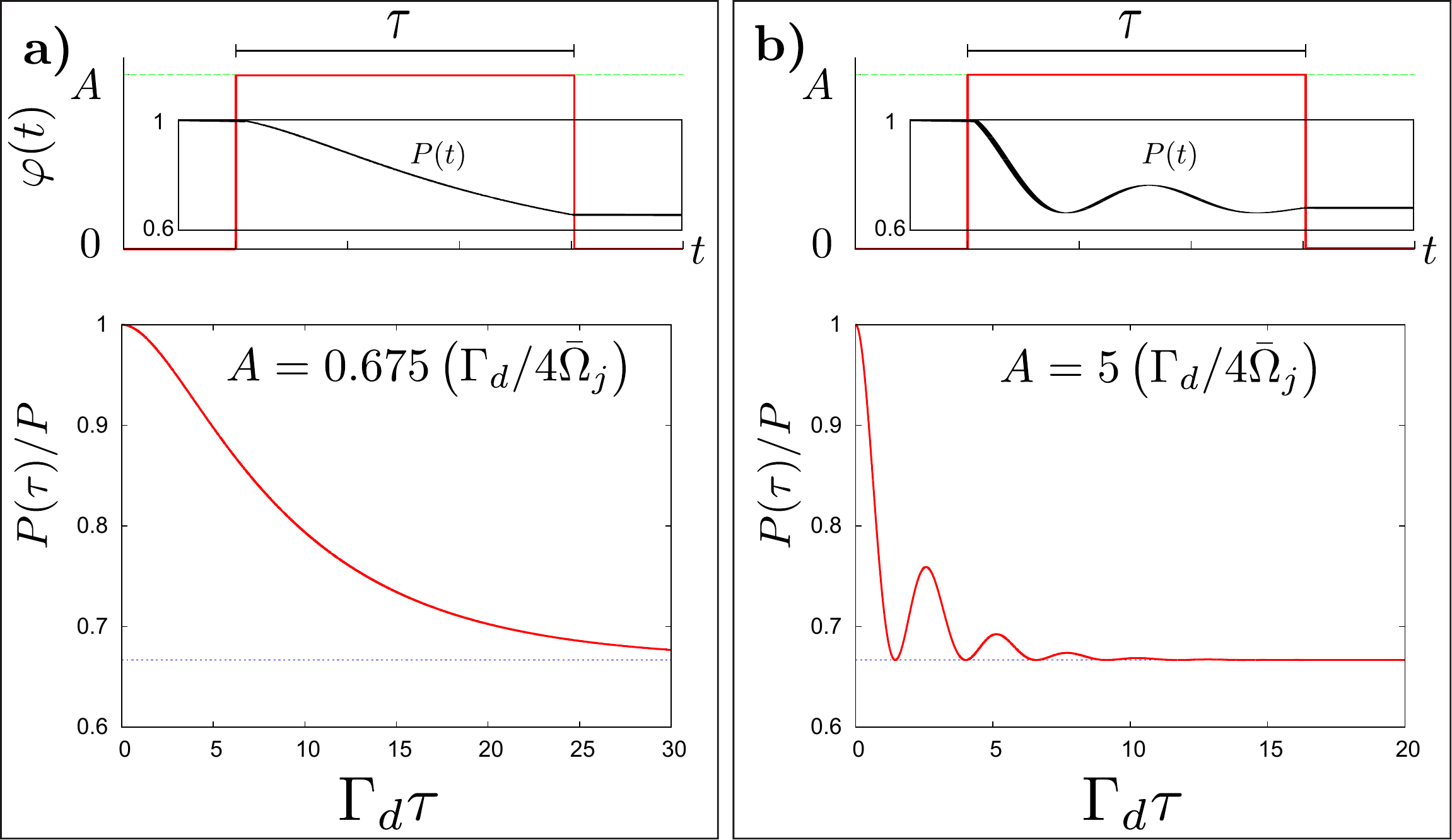}
\caption{Singlet-to-triplet manipulation. The upper plots show the pulse $\varphi(t)$ of duration $\tau$; the insets show the time-dependence of the probability $P$ to realize the long-lived triplet state. The lower plots depict the dependence $P(\tau)$, in \textbf{a)} for the case $A<\Gamma_d/4\bar{\Omega}_j$, and in \textbf{b)} for the case $A>\Gamma_d/4\bar{\Omega}_j$.}
\label{fig:initialization}
\end{figure}

\begin{figure}
\onefigure[width=0.8\linewidth]{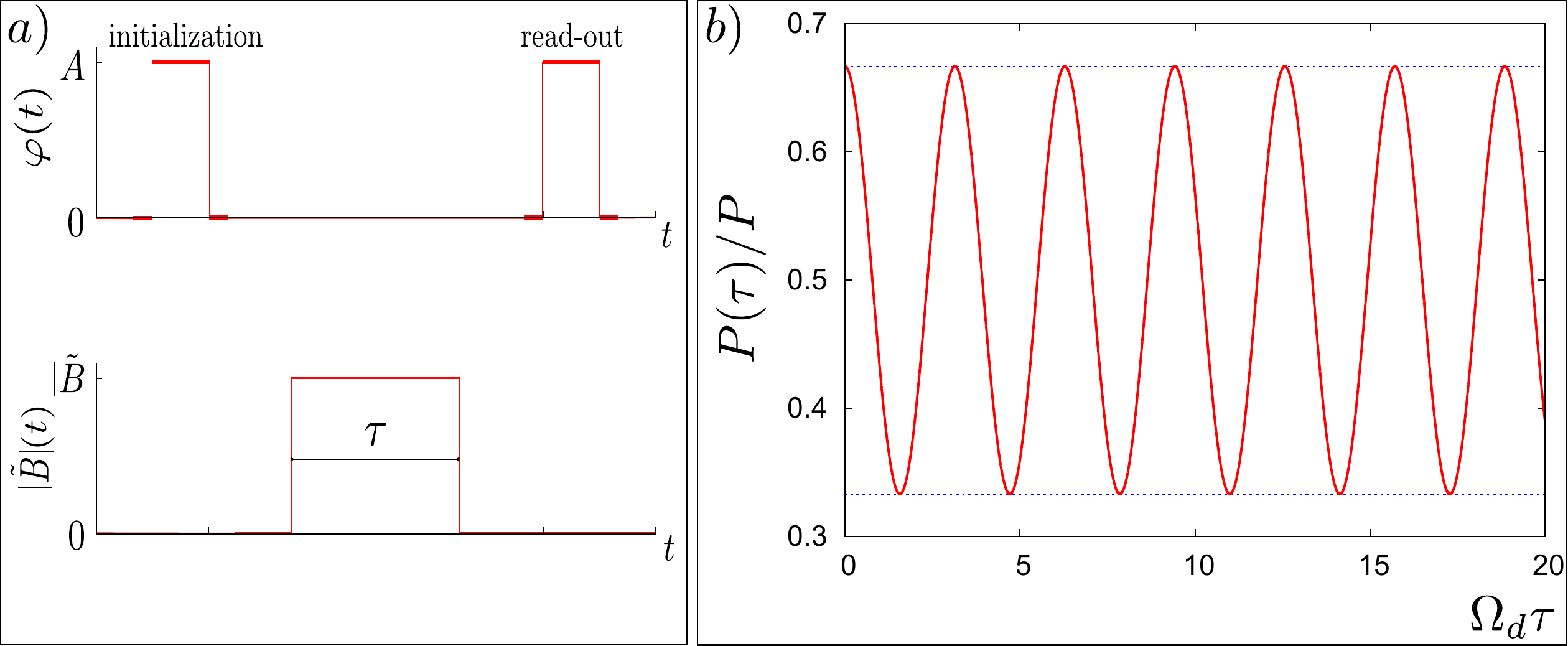}
\caption{Qubit initialization, rotation and read-out. \textbf{a)} The three square pulses are applied in succession. The upper plot shows the two pulses of $\varphi(t)$ used for initialization and read-out. The duration of the pulses is optimized to transfer all population from state $\ket{0}$ to $\ket{S}$. The lower plot shows the magnetic field pulse of duration $\tau$ used for triplet-to-triplet manipulation. \textbf{b)} The dependence $P(\tau)$ of the probability to measure the lower critical current after the sequence of three pulses.}
\label{fig:rotation}
\end{figure}

Let us give an example of an experiment with the qubit. In the beginning, the probabilities of all triplet states are the same, $p_0=p_1=p_2=P/3$, $P$ being the probability to realize a long-lived triplet state introduced above. These probabilities can be manipulated by triplet-to-singlet transitions. By tuning the pulse duration (Fig.~\ref{fig:initialization}), we can achieve the full depletion of $p_0$ {\it initializing} the qubit to a mixed state with $p_0=0,p_1=p_2=P/3$.  Subsequent {\it rotation} works if the qubit is in $\ket{1}$ bringing it to a superposition of states $\ket{0}$ and $\ket{1}$ by a triplet-to-triplet pulse of variable duration. Subsequently we {\it read} the resulting $p_0$ after the rotation  using the same pulse as used for the initialization. Fig.~\ref{fig:rotation} shows the result of applying the pulses for initialization, qubit rotation, and read-out in succession. The measured probability to get the lower critical current $P(\tau) = 2P/3 -p_0(\tau)$, is plotted versus the pulse duration $\tau$ manifesting Rabi oscillations.

%%See fig.~\ref{fig.1}, table~\ref{tab.1} and eq.~(\ref{eq.1}).
%%See also~\cite{b.a,b.b}.
%%\begin{equation}
%%\label{eq.1}
%%0\neq1
%%\end{equation}

\section{Conclusion} We have presented a proposal of a novel qubit design using the spin states of two superconducting quasiparticles trapped in a junction. Read-out of the qubit is based on spin-blockade that inhibits recombination of quasiparticles in the triplet state. We have described the resonant manipulation of singlet-to-triplet and triplet-to-triplet transitions and have explained the operation of the qubit. The qubit operation frequency estimated as $\Omega\simeq 10^7$ Hz is much larger than the qubit relaxation rate $\Gamma$, $\Omega/\Gamma\simeq 10^7$. Realization of our proposal would unambiguously demonstrate for the first time the spin properties of superconducting quasiparticles.

\acknowledgments
This work is part of the research program of the Stichting FOM.

\end{document}